\documentstyle[twoside,fleqn,espcrc2,fps]{article}

\newcommand{\AmS}{{\protect\the\textfont2
  A\kern-.1667em\lower.5ex\hbox{M}\kern-.125emS}}

\title{The Schwinger Model with Perfect Staggered Fermions}

\author{W. Bietenholz\address{HLRZ c/o Forschungszentrum J\"{u}lich,
D-52435 J\"{u}lich, Germany}
\thanks{Present address: NORDITA,
Blegdamsvej 17, DK-2100 Copenhagen O$\!\!\!\!\!\, /$ , Denmark.}
        and 
H. Dilger\address{Institute for Theoretical Physics I, WWU M\"{u}nster,
Wilhelm-Klemm Str. 9, D-48149 M\"{u}nster, Germany}
\thanks{Poster presented at LAT98.}
}
\begin{document}

\begin{abstract}
We construct a new perfect action for free
staggered fermions, which is more local
than the one obtained from the standard block average
scheme. This pays off in superior properties after a 
short ranged truncation.
This action is ``gauged by hand'' and
tested in Schwinger model simulations
by means of a new variant of hybrid MC.
Using ``fat links'' for the gauge field, we obtain a tiny
``pion'' mass down to
$\beta $~\raisebox{-.4ex}{$\stackrel{<}{\sim}$}~1.5, 
and the ``eta'' mass
follows very closely the prediction of asymptotic scaling.
\end{abstract}

\maketitle

It has been observed that a good implementation of classically
perfect actions provides an excellent level of improvement in
the sense that lattice artifacts are drastically suppressed.
However, the most successful applications have been limited
to two dimensions so far, and they involve a huge number
of couplings \cite{cp2d,lang}. This implies that
corresponding 4d actions are hardly
applicable. Therefore it is crucial to achieve good improvement
with {\em rather few extra terms}, and also this can be studied first
in $d=2$. However, the known classically perfect actions of Refs.
\cite{cp2d,lang} are not 
local enough for a very short ranged truncation to be sensible,
hence -- in view of $d=4$ -- different construction 
methods should be considered.
Here we present a successful application of ``gauging by hand'', 
i.e. we use the truncated perfect couplings of the free fermion
and insert gauge couplings by hand. This leads to a relatively
modest overhead in the number of couplings compared to the standard
action, but to an improvement on the same level as classical
perfection. We conclude that this procedure is promising for $d=4$.\\

Perfect actions are constructed by iterating block variable
renormalization group transformations (RGTs). For massless
staggered fermions, the RGT should not mix the (pseudo-)flavors 
in order to preserve the remnant chiral symmetry 
$U(1) \otimes U(1)$. This can be achieved by blocking each
flavor separately \cite{kalk}. As an improvement over the
standard block average (BA) scheme \cite{perfstag}, we propose
``{\bf partial decimation}'' (PD) \cite{pardec}. The free propagator
$\Delta_{BA}$ is built by averaging source and sink in each block,
whereas in the PD scheme one only averages the source and 
treats the sink by decimation, or vice versa.
Both lead to local perfect actions for free staggered fermions,
and in both cases we optimize the locality by tuning the RGT
parameters. The optimization criterion is that the mapping down to
$d=1$, $\Delta^{-1}(p_{1},0,\dots ,0)$,
only  couples nearest neighbors. For the PD scheme
this is the case for a simple $\delta $ function blocking at $m=0$
\cite{pardec}. However, that criterion can be fulfilled
at arbitrary mass $m$.
In $d>1$ the PD scheme leads to a superior degree of locality,
i.e.~to a faster exponential decay of the couplings.
Hence the truncation can be expected to be less harmful.

The truncation itself should also be done with some care.
We found {\bf mixed periodic boundary conditions} to be optimal:
we impose {\em anti-periodicity} over 6 lattice spacings in the direction
of odd coupling distance, and {\em periodicity} over 6 lattice spacings
in the other direction(s). The resulting couplings in $d=2$
are given in the following Table; 
for $d=4$ we refer again to Ref. \cite{pardec}.
There we also discuss massive fermions, including the case
of non-degenerated flavors.

\begin{center}
\begin{tabular}{|c|c|c|}
\hline
$m=0$ & BA & PD \\
\hline
\hline
$c_{1,0}$ &  ~0.866299 & {\bf ~0.878234}\\
\hline
$c_{1,2}$ & ~0.066850 & {\bf ~0.060883}\\
\hline
$c_{3,0}$ &  ~0.016043 & {\bf ~0.010425}\\
\hline
$c_{3,2}$ & -0.008022 &{\bf -0.005212} \\
\hline
\end{tabular}
\vspace*{3mm}
\end{center}

The better quality of the truncated perfect action arising from the
PD scheme is confirmed by the dispersion relation and by the
thermodynamic scaling behavior \cite{pardec,bary}.
In the latter case, the scaling region is extended by one order 
of magnitude compared to the standard action.\\

In our {\bf application to the Schwinger model} we obtain
a ``$\pi$'' and a ``$\eta$'' particle, since we deal with
two massless flavors. We use the couplings of the PD scheme
in the above Table, and we include the following lattice paths:


\begin{figure}[hbt]
\vspace{-6mm}
\def\fpsangle{270}
\epsfxsize=37mm
\fpsbox{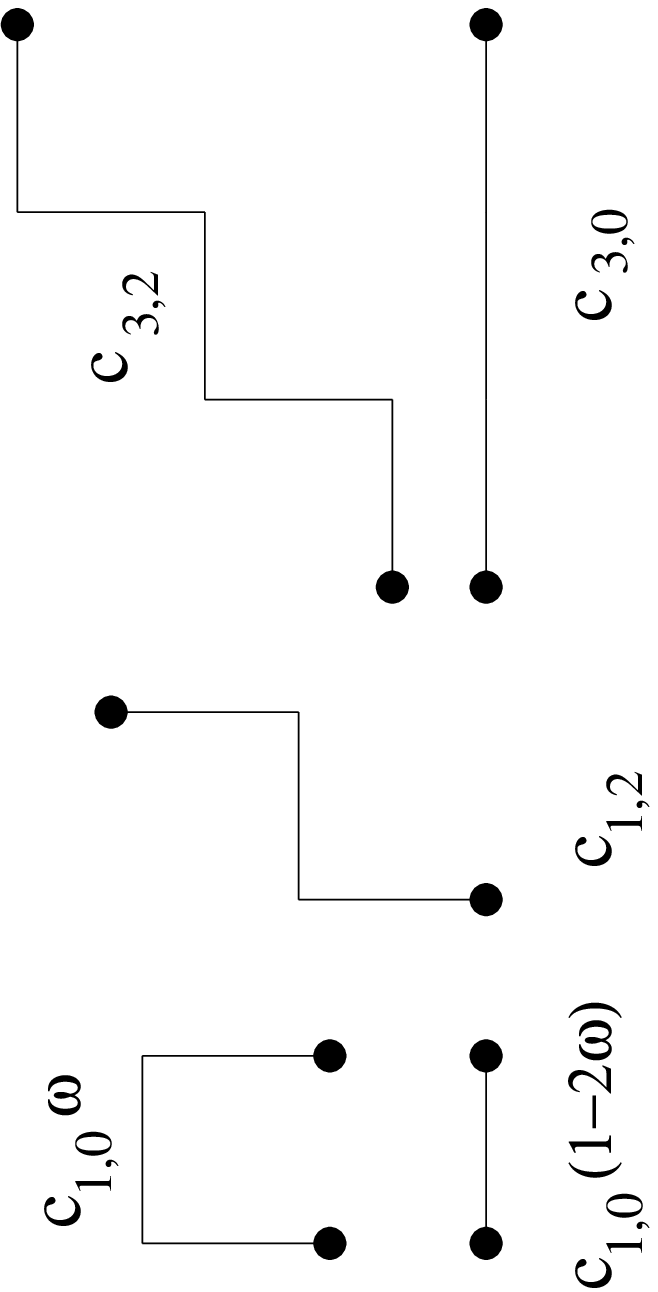}
\vspace{-7mm}
\end{figure}

The staple weight $\omega$ is tuned so that the smallest
absolute values of fermion matrix eigenvalues are minimized.
We find ${\bf \omega =0.238}$ to be optimal, 
and we use the resulting fat link.

For the pure gauge part, we use an ultralocal plaquette action, 
which is perfect in $d=2$ \cite{quaglu,prep}.
 
In our action, a fixed site couples to 24 other sites, which is
an overhead of a factor of 6 compared to the standard action.
In view of QCD, we design an algorithm which avoids a proportional
increase in compute time:\\

{\em Variant of Hybrid Monte Carlo}\\
(1) Possible Molecular Dynamics steps are identified using the
standard action.\\
(2) The acceptance decision is based on the quasi-perfect action.\\

The crucial question is the acceptance rate; it is good down to a
minimal value of $\beta$. If we include the fat link in (1), then it
works well down to very small $\beta$ 
$(\beta $~\raisebox{-.4ex}{$\stackrel{<}{\sim}$}~1.5).
Still the simplification in (1) accelerates the algorithm by a factor
of $\approx 4$, while the acceptance rate is divided by $\approx 2$.

We generate $3 \cdot 10^{5}$ configurations 
on a {\bf ${\bf 16\times 16}$ lattice}, and we measure after
multi-steps of 100 configurations
\footnote{The number 100 corresponds to the maximal autocorrelation
time for topological observables (for other quantities it is 
$\approx 25$).}.
The simulation results \cite{prep} are unquenched.\\

The {\bf spectrum} is determined from local bilinears \cite{dilg}:
\begin{eqnarray*} && \hspace{-7mm}
M^{1-}_{--}(x) = (-1)^{x_{1}+x_{2}}[\bar \psi_{x} \psi_{x+(1,0)}
- \bar \psi_{x+(1,0)} \psi_{x}] \\ && \hspace{-7mm}
M^{1+}_{++}(x) = \bar \psi_{x} \psi_{x+(1,0)}
+ \bar \psi_{x+(1,0)} \psi_{x} \ .
\end{eqnarray*}
We measure the correlators \
$C^{\alpha}_{\sigma_{1} \sigma_{2}}(p_{1},x_{2}) =$
\begin{displaymath}
\sum_{x_{1}} \langle M^{\alpha}_{\sigma_{1}\sigma_{2}}(x_{1},x_{2})
M^{\alpha}_{\sigma_{1}\sigma_{2}}(0,0) \rangle e^{ix_{1}p_{1}} \ .
\end{displaymath}
\begin{itemize}
\item The isovector current correlation is obtained from 
$C^{1-}_{--}$. 
It has only connected contributions and yields the ``pion''.

\item  The isoscalar current correlation is obtained from
$C^{1+}_{++}$. 
It has disconnected contributions too and yields the ``eta particle''.
\end{itemize}

At $\beta =3$ we evaluate the dispersion relation
$E^{2}(p_{1})=m^{2}+p_{1}^{2}$ from cosh fits for $m_{\pi}$ and
$m_{\eta}$. The result is shown in Fig. 1.
Then we measure these masses with the standard action
(from $p_{1}=0$) and with the quasi-perfect action (from $p_{1}=0$
and $p_{1}= \pi /8$), see Fig. 2. The continuum predictions are
$m_{\pi}=0$ ({\bf scaling}), $m^{2}_{\eta}=2/(\pi \beta )$
({\bf asymptotic scaling}).\\

\begin{figure}[hbt]
\def\fpsangle{270}
\epsfxsize=57mm
\fpsbox{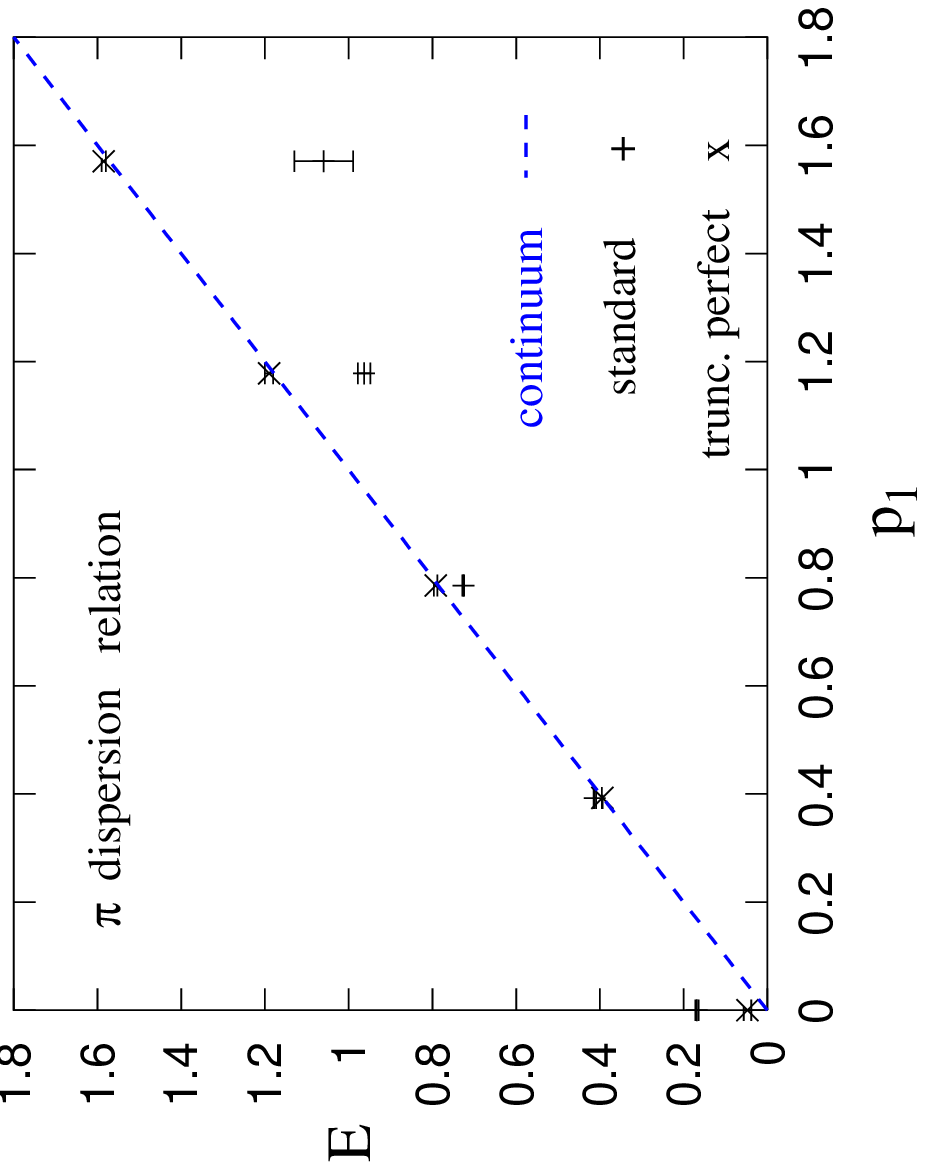}
\vspace{2mm}
\epsfxsize=55mm
\fpsbox{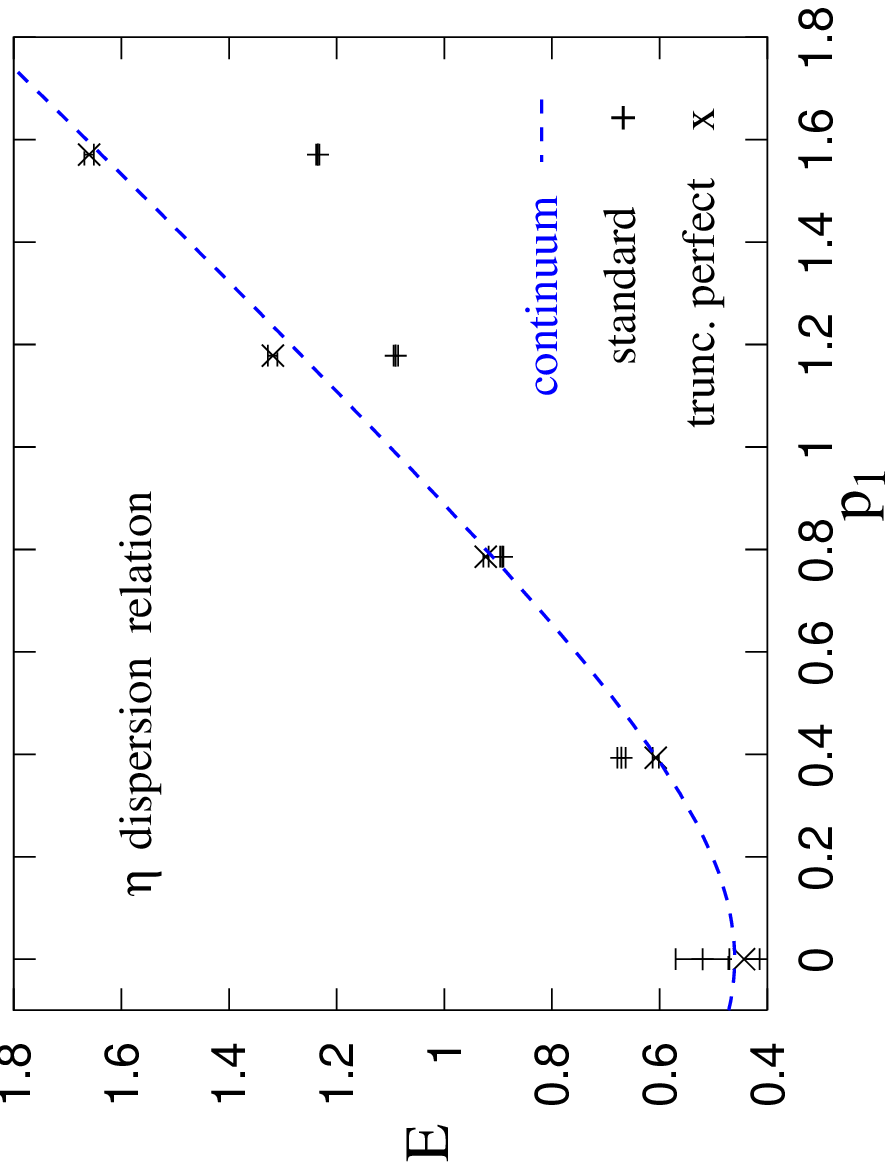}
\vspace{-7mm}
\caption{``Meson'' dispersion relations at $\beta=3$.}
\vspace{-7mm}
\end{figure}

\begin{figure}[hbt]
\def\fpsangle{0}
\epsfxsize=68mm
\fpsbox{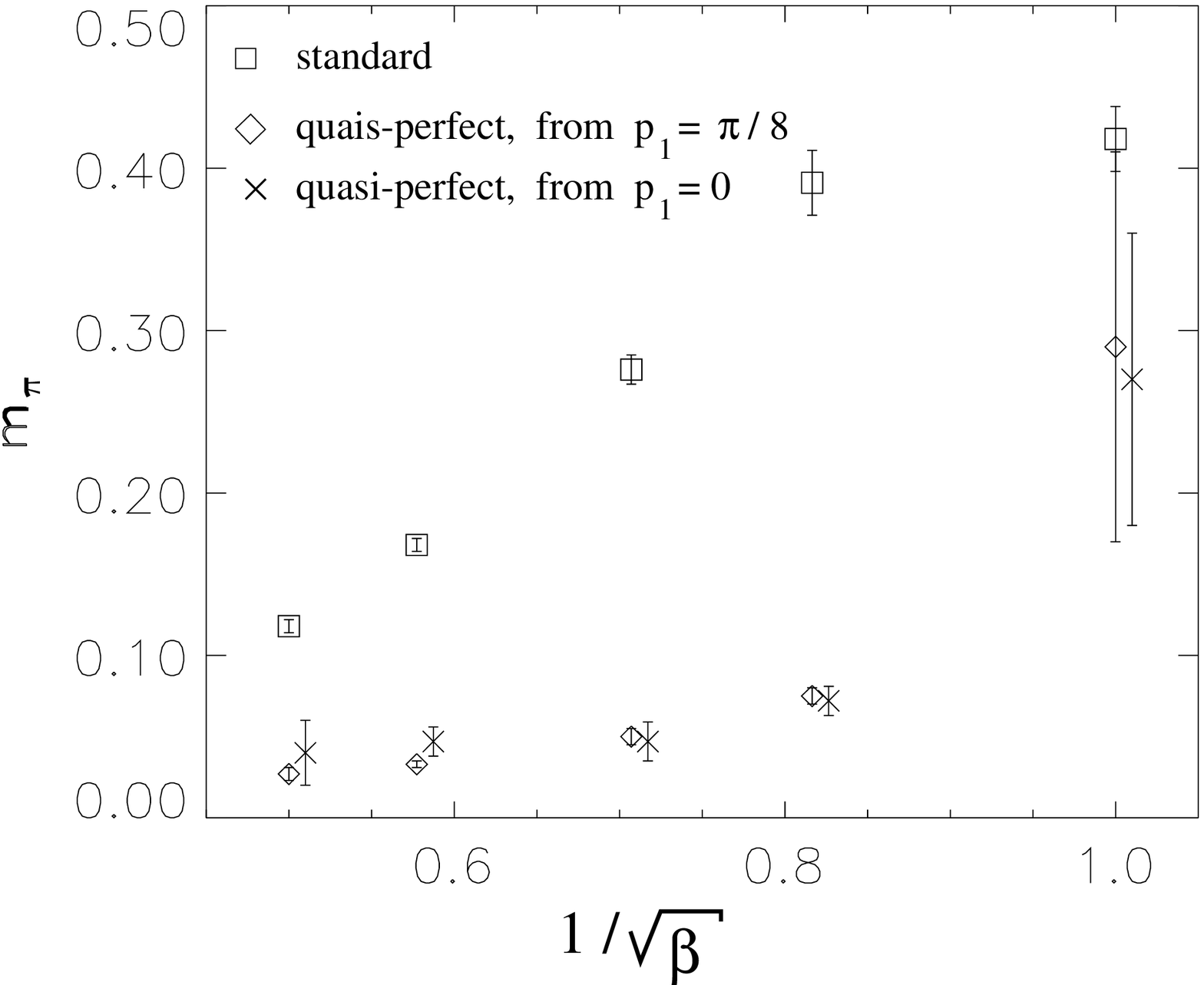}
\epsfxsize=68mm
\fpsbox{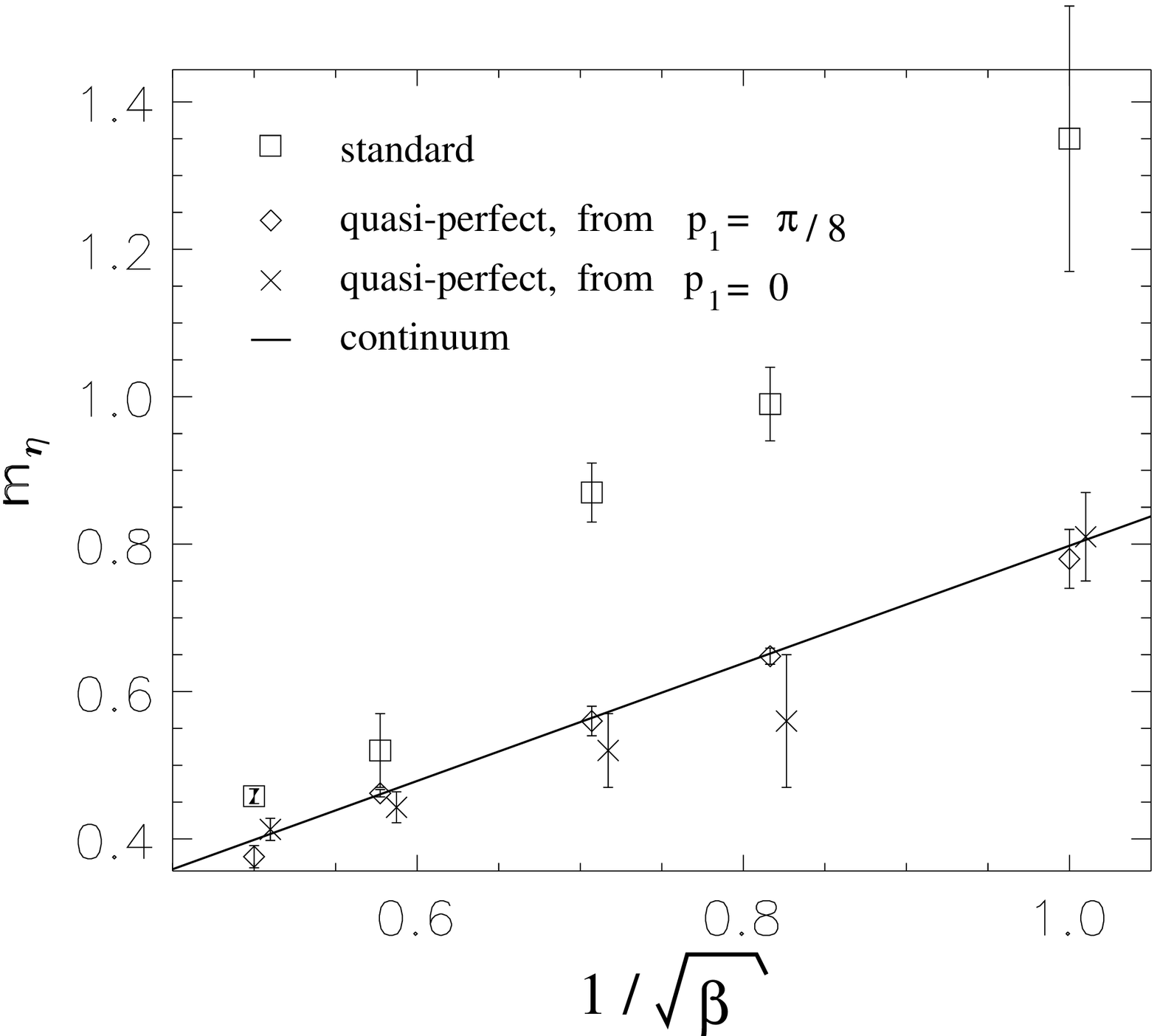}
\vspace{-8mm}
\caption{The masses $m_{\pi}$ and $m_{\eta}$ measured
at various values of $\beta$.}
\vspace{-7mm}
\end{figure}

Using the quasi-perfect action described above,
we find a very small $m_{\pi}$, i.e.~very good scaling,
down to 
$\beta $~\raisebox{-.4ex}{$\stackrel{<}{\sim}$}~1.5.
The $\eta$ dispersion
follows very closely the prediction of asymptotic scaling
down to $\beta \approx 1$.
These results ---
in particular $m_{\pi}$, which tests the level of improvement --- 
are even better than those of Ref. \cite{lang},
which used truncated perfect Wilson fermions together with a
classically perfect vertex function (parameterized by 123
independent couplings).

This shows that a truncated perfect free fermion, which is suitably
``gauged by hand'', {\em can} represent a highly improved, short-ranged
action. This program is applicable and promising for QCD \cite{HF}.


\vspace*{-2mm}
\begin{thebibliography}{20}
\vspace*{-1mm}

\bibitem{cp2d} P. Hasenfratz and F. Niedermayer,
Nucl. Phys. B 414 (1994) 785. 
R. Burkhalter, Phys. Rev. D 54 (1996) 4121.

\bibitem{lang} C. B. Lang and T. K. Pany, Nucl. Phys. B 513 (1998) 645.

\bibitem{kalk} T. Kalkreuter, G. Mack and M. Speh,
Int. J. Mod. Phys. C 3 (1992) 121.

\bibitem{perfstag} 
W. Bietenholz, E. Focht and U.-J. Wiese, Nucl. Phys. B 436 (1995) 385.
H. Dilger, Nucl. Phys. B 490 (1997) 331.
W. Bietenholz, R. Brower, S. Chandrasekharan and U.-J. Wiese,
Nucl. Phys. B 495 (1997) 285.

\bibitem{pardec} W. Bietenholz and H. Dilger, \\
hep-lat/9803018.

\bibitem{bary} W. Bietenholz, hep-lat/9805014.

\bibitem{quaglu} W. Bietenholz and U.-J. Wiese, Nucl. Phys. B 
464 (1996) 319.

\bibitem{prep} W. Bietenholz and H. Dilger, in prep.


\bibitem{dilg} H. Dilger, Nucl. Phys. B 434 (1995) 321;
Int. J. Mod. Phys. C 6 (1995) 123.


\bibitem{HF} W. Bietenholz et al.,
Nucl. Phys. B (Proc. Suppl.) 53 (1997) 921.
K. Orginos et al., 
Nucl. Phys. B (Proc. Suppl.) 60 (1998) 904.
T. DeGrand, hep-lat/9802012.
W. Bietenholz et al., hep-lat/9807013 and talk 
by N. Eicker at LAT98.


\end{thebibliography}
\end{document}